\documentclass[aps,pra,twocolumn,superscriptaddress,floatfix]{revtex4}

\usepackage{amssymb}
\usepackage{graphicx}
\usepackage[dvips]{epsfig}
\usepackage{amsmath}
\usepackage{wrapfig}
\usepackage{color}
\usepackage{subfig}
\usepackage{enumitem}
\usepackage{epstopdf}
\usepackage{placeins}

\expandafter\let\csname equation*\endcsname\relax
\expandafter\let\csname endequation*\endcsname\relax

\newcommand{\hilight}[1]{\colorbox{green}{#1}} 
\newcommand{\braket}[3]{\left\langle  #1  || #2 || #3 \right\rangle}

\renewcommand{\thefigure}{\Roman{figure}\cite{lacroute2012}}

\begin{document}
\title{Corrections to our results for optical nanofiber traps in Cesium}
\author{D. Ding}
\address{Norman Bridge Laboratory of Physics, California Institute of Technology} 
\author{A. Goban}
\address{Norman Bridge Laboratory of Physics, California Institute of Technology}
\author{K. S. Choi}
\address{Norman Bridge Laboratory of Physics, California Institute of Technology}
\address{Spin Convergence Research Center, Korea Institute of Science and Technology}
\author{H. J. Kimble}
\address{Norman Bridge Laboratory of Physics, California Institute of Technology}

\begin{abstract}
Several errors in Refs. \cite{lacroute2012,goban2012} are corrected related to the optical trapping potentials for a state-insensitive, compensated nanofiber trap for the $D_2$ transition of atomic Cesium. Section I corrects our basic formalism in Ref. \cite{lacroute2012} for calculating dipole-force potentials. Section II corrects erroneous values for a partial lifetime and a transition wavelength in Ref. \cite{lacroute2012}. Sections III and IV present corrected figures for various trapping configurations considered in Refs. \cite{lacroute2012} and \cite{goban2012}, respectively.

\end{abstract}

\date{\small\today}

\maketitle
\pagestyle{myheadings}
\markboth{}{Corrections to our results for optical nanofiber traps in Cesium}
\thispagestyle{empty}

\section{Formalism}
\label{formal}

\vspace{-4mm}

The light shifts calculated in Ref. \cite{lacroute2012} are based upon Eq. (2) in Ref. \cite{lacroute2012}. The text states that the basis for Eq. (2) is `spherical' (i.e., irreducible spherical tensors). In fact, the basis for Eq. (2) is actually a Cartesian basis (i.e., $x,y,z$), which we used to perform all calculations in Ref. \cite{lacroute2012}. 

In Eq. (2) of Ref. \cite{lacroute2012}, $\alpha^{(0)}$,$\alpha^{(1)}$ and $\alpha^{(2)}$ are the scalar, vector and tensor atomic dynamic polarizabilities defined in Ref. \cite{deutsch}. The dipole matrix element in Ref. \cite{lacroute2012} is $d_{JJ'}^2=|\braket{J}{\mathbf{d}}{J'}|^2=\frac{3\pi\epsilon_0\hbar c^3}{\omega_{J'J}^3}\frac{2J'+1}{2J+1}\frac{1}{\tau_{J'J}}$, as defined in Ref. \cite{dstecktext}. However, as Ref. \cite{deutsch} only considered light shifts of the ground states, the definition for $d_{JJ'}$ in Ref. \cite{dstecktext} becomes problematic when defining $d_{JJ'}$ for the excited states. $J$ and $J'$ are defined with respect to states of lower to higher energy, respectively, rather than with respect to initial and final states. This notation is therefore ambiguous if the initial state is an excited state. Also, the counter-rotating term was not taken into account in Ref. \cite{deutsch}. Therefore, we follow the formalism in Refs. \cite{rosenbusch, Beloy2009,fam2012}, in which case Eq. (2) of Ref. \cite{lacroute2012} should be rewritten as follows:
\begin{widetext}
\begin{equation}
\begin{array}{llll}
\hat{H}_{\mathrm{ls}}&=-\alpha^{(0)}\mathbf{\hat{E}^{(-)}}\cdot \mathbf{\hat{E}^{(+)}}-i\alpha^{(1)}\frac{\left(\mathbf{\hat{E}^{(-)}}\times \mathbf{\hat{E}^{(+)}} \right) \cdot \mathbf{\hat{F}}}{2F}-\sum\limits_{\mu,\nu}\alpha^{(2)}\hat{E}^{(-)}_{\mu}\hat{E}^{(+)}_{\nu}\frac{3}{F(2F-1)}\left [\frac{1}{2}(\hat{F}_{\mu}\hat{F}_{\nu}+\hat{F}_{\nu}\hat{F}_{\mu})-\frac{1}{3}\hat{F}^2\delta_{\mu\nu}\right],
\end{array}
\label{hamiltonian}
\end{equation} 
\end{widetext}
 where the dipole matrix element in  Refs. \cite{rosenbusch, Beloy2009,fam2012} is $d_{JJ'}^2=|\braket{J}{\mathbf{d}}{J'}|^2=\frac{3\pi\epsilon_0\hbar c^3}{\omega_{J'J}^3}(2J'+1)\frac{1}{\tau_{J'J}}$, where $(J, J')$ are for (lower, upper) levels, respectively. $\alpha^{(0)}$, $\alpha^{(1)}$, and $\alpha^{(2)}$ of Eq. \eqref{hamiltonian} include counter-rotating terms. However, in Ref. \cite{lacroute2012}, we made an error in the definition of the vector polarizability, $\alpha^{(1)}$, by neglecting the rank dependence of the counter-rotating terms. In this errata, we correct the definition of $\alpha^{(1)}$ to incorporate the rank dependence of the counter-rotating terms in our calculations of light shifts \cite{rosenbusch, Beloy2009,fam2012}. 

The dynamic polarizabilities are then given by 
\begin{widetext}
\begin{eqnarray}
\alpha^{(0)}(J,F)&=&\displaystyle\sum\limits_{nJ'F'} \frac{d_{JJ'}^2}{3}(2F'+1)\begin{Bmatrix} 
 J    & J' & 1\\ 
 F'   & F & I 
 \end{Bmatrix}^2G^{(0)}_{FF'}\label{alpha0}
 \end{eqnarray}

\begin{eqnarray}
\alpha^{(1)}(J,F)&=&2\displaystyle\sum\limits_{nJ'F'} (-1)^{F+F'}d_{JJ'}^2 \sqrt{\frac{3F(2F+1)}{2(F+1)}}\begin{Bmatrix} 
 1    & 1 & 1\\ 
 F   & F & F' 
\end{Bmatrix}\notag\\
&&\times(2F'+1)\begin{Bmatrix} 
 J    & J' & 1\\ 
 F'   & F & I 
  \end{Bmatrix}^2 G^{(1)}_{FF'} \label{alpha1}\\
\alpha^{(2)}(J,F)&=&\displaystyle\sum\limits_{nJ'F'} (-1)^{F+F'} d_{JJ'}^2 \sqrt{\frac{10F(2F+1)(2F-1)}{3(F+1)(2F+3)}}\begin{Bmatrix} 
 1    & 1 & 2\\ 
 F   & F & F' 
 \end{Bmatrix} \notag\\
 &&\times(2F'+1)\begin{Bmatrix} 
 J    & J' & 1\\ 
 F'   & F & I 
  \end{Bmatrix}^2 G^{(2)}_{FF'} ,\label{alpha2}
\end{eqnarray}
\clearpage
\end{widetext}

where $G^{(K)}_{ij}$ is the rank-$K$ propagator defined as

\begin{eqnarray}
G^{(K)}_{ij}=\frac{1}{\hbar}\left\{\frac{1}{\omega_{ji}-\omega}+\frac{(-1)^K}{\omega_{ji}+\omega}\right\}.
\label{Gk}
\end{eqnarray}

We note that the counter-rotating term gains an overall minus sign for $K=1$ in the expression for the vector propagator $G_{FF^{\prime}}^{(1)}$. We have confirmed that our corrected formula for the dynamic polarizabilities (i.e., Eqs. (1-5)) is now consistent with the expressions in Ref. \cite{rosenbusch}.  

\vspace{-4mm}

\section{Partial Lifetimes and Transition Wavelengths}
\label{comp}

\vspace{-2mm}

In addition to corrections to our formalism discussed above, we correct two numerical errors for the atomic data used to calculate the light shifts in Refs. \cite{lacroute2012, goban2012}. These errors stem from a mistake for a partial lifetime $\tau$ and from an error for a transition wavelength $\lambda$. All values for $\tau$ and $\lambda$ of Ref. \cite{lacroute2012} were taken from Tables 7.1--7.3 of J. McKeever's Ph.D Thesis \cite{mckeever}. The specific errors are as follows: 

\begin{enumerate}
\item{The wavelength for the transition $7S_{1/2}\leftrightarrow6P_{3/2}$ in Table 7.2 of Ref. \cite{mckeever} is listed as $469.5$ nm, whereas the correct value is $1469.5$ nm.}
\item{The partial lifetime for the transition $7D_{3/2}\leftrightarrow6P_{3/2}$ in Table 7.3 of Ref. \cite{mckeever} is listed as $709.7$ $\mu$s, but instead should be $0.7097$ $\mu$s.}
\end{enumerate}
The corrected values for Tables 7.1--7.3 of Ref. \cite{mckeever} are highlighted with green color in Tables \ref{comparisontable1} and \ref{comparisontable2} of this errata. 

The partial lifetimes and wavelengths from Table 7.1--7.3 of Ref. \cite{mckeever} were obtained from Refs. \cite{laplanche,fabry} with the exception of the values for $6P\rightarrow6S$ from Ref. \cite{dsteck} and $6D\rightarrow6P$ from Ref. \cite{christina}. 

To further confirm the accuracy of the corrected values from Ref. \cite{mckeever}, we refer to Tables V and VI of Arora \textit{et al.} \cite{arora2007} (heretofore, labelled as `Clark'). Tables \ref{comparisontable1} and \ref{comparisontable2} in this errata compare the corrected tables from Ref. \cite{mckeever} to Tables V--VI of Ref. \cite{arora2007}. Our values agree well with those from Ref. \cite{arora2007}.

We note that all previous publications prior to Ref. \cite{lacroute2012} from our group, including Ref. \cite{mckeeverprl}, have used the correct values in Tables \ref{comparisontable1}--\ref{comparisontable2} and not those from McKeever's thesis \cite{mckeever}. Furthermore, these earlier results are for linearly polarized trapping fields for which the contribution of the vector term $\alpha^{(1)}$ vanishes. Therefore, these earlier results from our group stand without revision. 

Using the revised formalism from Section I and based on the corrected numerical values in Tables \ref{comparisontable1}--\ref{comparisontable2} from Section II, we present corrected figures to replace those in Refs. \cite{lacroute2012, goban2012} as described in the following sections.

\vspace{4mm}

\section{Corrected Figures for Ref. \cite{lacroute2012}}
\label{results}

\vspace{-2mm}

The figure numbering in this errata mirrors that of Ref.\cite{lacroute2012} with corrected figures given here in Roman numerals. Fig. \ref{fig6} relates to magic wavelengths for the Cs $D_2$ line and replaces Fig. 6 in Ref. \cite{lacroute2012}. The magic wavelength in this note is defined by the weighted average of the Zeeman sublevels $m_F$ (i.e., not by $\alpha^{(0)}$ alone).

As in Ref. \cite{lacroute2012}, we include a surface interaction potential of an atom with the dielectric nanofiber in our calculation of the total atomic trap potential. The surface potential of the ground state Cs atom near a planar dielectric surface can be approximated by the van der Waals potential $-\frac{C_3}{d^{3}}$, where $d=r-a$ and $C_3/\hbar=1.16~\text{kHz } \mu \text{m}^3$ \cite{lacroute2012,stern2011}.

In Figs.~\ref{fig7}, \ref{fig8}, and \ref{fig9}, the two-color evanescent trap from Ref. \cite{vetsch2010} is constructed from a pair of counter-propagating $x$-polarized ($\varphi_0 = 0$) red-detuned beams ($P_{\text{red}} = 2\times2.2$ mW) at $\lambda_{\text{red}} = 1064$ nm, forming an optical lattice, and a single repulsive $y$-polarized ($\varphi_0 = \pi/2$) blue-detuned beam ($P_{\text{blue}} = 25$ mW) at $\lambda_{\text{blue}} = 780$ nm. The SiO$_2$ tapered optical fiber has radius $a = 250$ nm in the trapping region. Figs. \ref{fig7},\ref{fig8}, \ref{fig9} replace figures of Fig. 7, 8, 9 of Ref. \cite{lacroute2012}. 

For the magic, compensated trap in Figs.~\ref{fig10}, \ref{fig11}, \ref{fig12}, we use a pair of counter-propagating $x$-polarized ($\varphi_0=0$) red-detuned beams ($P_{\text{red}}=2\times0.95$ mW) at the magic wavelength $\lambda_{\text{red}} = 935.3$ nm. Counter-propagating, $x$-polarized blue-detuned beams at a second magic wavelength $\lambda_{\text{blue}} = 684.9$ nm are used with a power $P_{\text{blue}} = 2\times16$ mW. The resulting interference is averaged out by detuning the beams to $\delta_{fb} = 30$ GHz. Figs. \ref{fig10}, \ref{fig11}, \ref{fig12} replace Figs. 10, 11, 12 of Ref. \cite{lacroute2012}. 

\vspace{-4mm}

\section{Corrected Figures for Ref. \cite{goban2012}}
\label{results}

\vspace{-2mm}

Because of the errors described in Sections I and II, our experiment in Ref. \cite{goban2012} used red- and blue-detuned beams at wavelengths $\lambda_{\text{red}} = 937.1$ nm  and $\lambda_{\text{blue}} = 686.1$ nm instead of the correct values of $\lambda_{\text{red}} \simeq 935.7$ nm and $\lambda_{\text{blue}} \simeq 684.8$ nm calculated in the same fashion as Fig. \ref{fig6} but now for $F=4$ of $6S_{1/2}$ to $F'=5$ of $6P_{3/2}$.
Fig. \ref{fig13} shows the trapping potentials for the ground and excited states for the correct magic wavelengths of $\lambda_{\text{red}} \simeq 935.7$ nm and $\lambda_{\text{blue}} \simeq 684.8$ nm for this transition. 

For the actual wavelengths $\lambda_{\text{red}} = 937.1$ nm and $\lambda_{\text{blue}} = 686.1$ nm used in our experiment \cite{goban2012}, Fig.1 and Fig. SM5 of Ref. \cite{goban2012} are here replaced by Figs. \ref{fig14} and \ref{fig15}, respectively, which incorporate the revisions described in Sections I and II.


\section{Conclusion}
\label{conclude}

\vspace{-3mm}

Our emphasis has been to correct the formalism (Section I) and atomic data (Section II and Tables I, II) that are the basis for our calculations in Refs. \cite{lacroute2012,goban2012}. Recently a more extensive set of atomic data than in Tables I and II has become available \cite{fam2012}. We have confirmed that these data with our formalism in Eqs. (1-5) reproduce Figs. (4, 5) from Ref. \cite{fam2012}.

However, the expanded set of atomic levels and lifetimes in Ref. \cite{fam2012} lead to small differences between Figs. (4, 5) \cite{fam2012} and corresponding figures computed from our Tables I, II. These differences are most pronounced around $685$nm (e.g., our Fig. VI(a)) principally due to excited-state contributions up to $n \sim 25$, which are not included in Tables I, II. We therefore recommend that the data set from Ref. \cite{fam2012} be employed for the calculation of ac Stark shifts for the $D_2$ line in atomic Cesium rather than the less extensive data in our Tables I, II.

\vspace{-4mm}

\begin{acknowledgments}

\vspace{-2mm}

We thank Dr. F. Le Kien and Prof. A. Rauschenbeutel for contacting us directly to alert us to the errors in Ref. \cite{lacroute2012} and for subsequent critical correspondence.
\end{acknowledgments}

\vspace{-3mm}

\FloatBarrier

\newpage

\FloatBarrier
\begin{table*}[h]
\begin{ruledtabular}
\begin{tabular}{lcrrr}
\multicolumn{1}{c}{Level $nS$} &
\multicolumn{1}{c}{$\lambda_{\text{McKeever}}$} &
\multicolumn{1}{c}{$\lambda_{\text{Clark}}$}&
\multicolumn{1}{c}{$\tau_{\text{McKeever}}$}&
\multicolumn{1}{c}{$\tau_{\text{Clark}}$} \\
\hline
	&(nm)	& (nm)	& ($\mu$s)	&($\mu$s)\\
\hline
6	&852.4	&852.35	&0.03051	&0.0306\\
7	&\hilight{1469.5}	&1469.89	&0.07529	&0.0749\\
8	&794.4	&794.61	&0.2599	&0.2319\\
9	&658.8	&658.83	&0.5533	&0.4759\\
10	&603.4	&603.58	&0.9924	&0.8374\\
11	&574.6		&&1.607	&\\
12	&557.3		&&2.428	&\\
13	&546.3		&&3.49	&\\
14	&538.5		&&4.809	&\\
15	&532.9		&&6.431	&

\end{tabular}   
\end{ruledtabular}
\caption{\label{comparisontable1} Comparison between corrected partial lifetimes in Ref. \cite{mckeever} with Ref. \cite{arora2007} for $6P_{3/2} \rightarrow 6S_{1/2}$~\cite{dsteck} and $nS_{1/2} \rightarrow 6P_{3/2}$ for $n=7-15$~\cite{laplanche}. Partial lifetime ($\tau_{\text{McKeever}}$) and wavelength ($\lambda_{\text{McKeever}}$) are from Ref. \cite{mckeever} with the corrected value highlighted in green. $\tau_{\text{Clark}}$ and $\lambda_{\text{Clark}}$ are from Ref. \cite{arora2007}. The conversion from $d$ to $\tau$ from Ref. \cite{arora2007} is done using Eq. (7.8) in Ref. \cite{mckeever}.}
\end{table*}

\begin{table*}[h]
\begin{ruledtabular}
\begin{tabular}{lccccrrrr}
\multicolumn{1}{c}{Level $nD$} &
\multicolumn{1}{c}{$\lambda_{3/2 ~\text{McKeever}}$}&
\multicolumn{1}{c}{$\lambda_{3/2 ~\text{Clark}}$} &
\multicolumn{1}{c}{$\tau_{3/2 ~\text{McKeever}}$}&
\multicolumn{1}{c}{$\tau_{3/2 ~\text{Clark}}$}&
\multicolumn{1}{c}{$\lambda_{5/2 ~\text{McKeever}}$} &
\multicolumn{1}{c}{$\lambda_{5/2 ~\text{Clark}}$} &
\multicolumn{1}{c}{$\tau_{5/2 ~\text{McKeever}}$}&
\multicolumn{1}{c}{$\tau_{5/2 ~\text{Clark}}$}\\
\hline
&(nm)	&(nm)	&($\mu$s)	&($\mu$s)	&(nm)	&(nm)&($\mu$s)	&($\mu$s)\\
\hline
5	&3612.7	&3614.09	&10.09	&9.2931	&3489.2	&3490.97	&1.433	&1.3692\\
6	&921.1	&921.11	&0.3466	&0.3497	&917.2	&917.48	&0.0587	&0.0604\\
7	&698.3	&698.54	&\hilight{0.7097}	&0.7061	&697.3	&697.52	&0.1198	&0.1203\\
8	&621.7	&621.93	&1.284	&1.2884	&621.3	&621.48	&0.217	&0.1566\\
9	&584.7	&		&2.131	&		&584.5	&		&0.3587	&\\
10	&563.7	&		&3.29	&		&563.5	&		&0.5527	&\\
11	&550.4	&		&4.807	&		&550.3	&		&0.8063	&

\end{tabular}   
\end{ruledtabular}
\caption{\label{comparisontable2} Comparison between corrected partial lifetimes in Ref. \cite{mckeever} with Ref. \cite{arora2007} for $nD_{(3/2,5/2)} \rightarrow 6P_{3/2}$. Partial lifetime ($\tau_{(3/2,5/2)~\text{McKeever}}$) and wavelength ($\lambda_{(3/2,5/2)~\text{McKeever}}$) are from Ref. \cite{mckeever} with the corrected value highlighted in green. $\tau_{(3/2,5/2)~\text{Clark}}$ and $\lambda_{(3/2,5/2)~\text{Clark}}$ are from Ref. \cite{arora2007}. The conversion from $d$ to $\tau_{(3/2,5/2)}$ from Ref. \cite{arora2007} is done using Eq. (7.8) in Ref. \cite{mckeever}.}
\end{table*}

\addtocounter{figure}{5}

\begin{figure*}[htbp]
\vspace{-5mm}
\centering
\label{fig6:subfig1}\includegraphics[width=8cm]{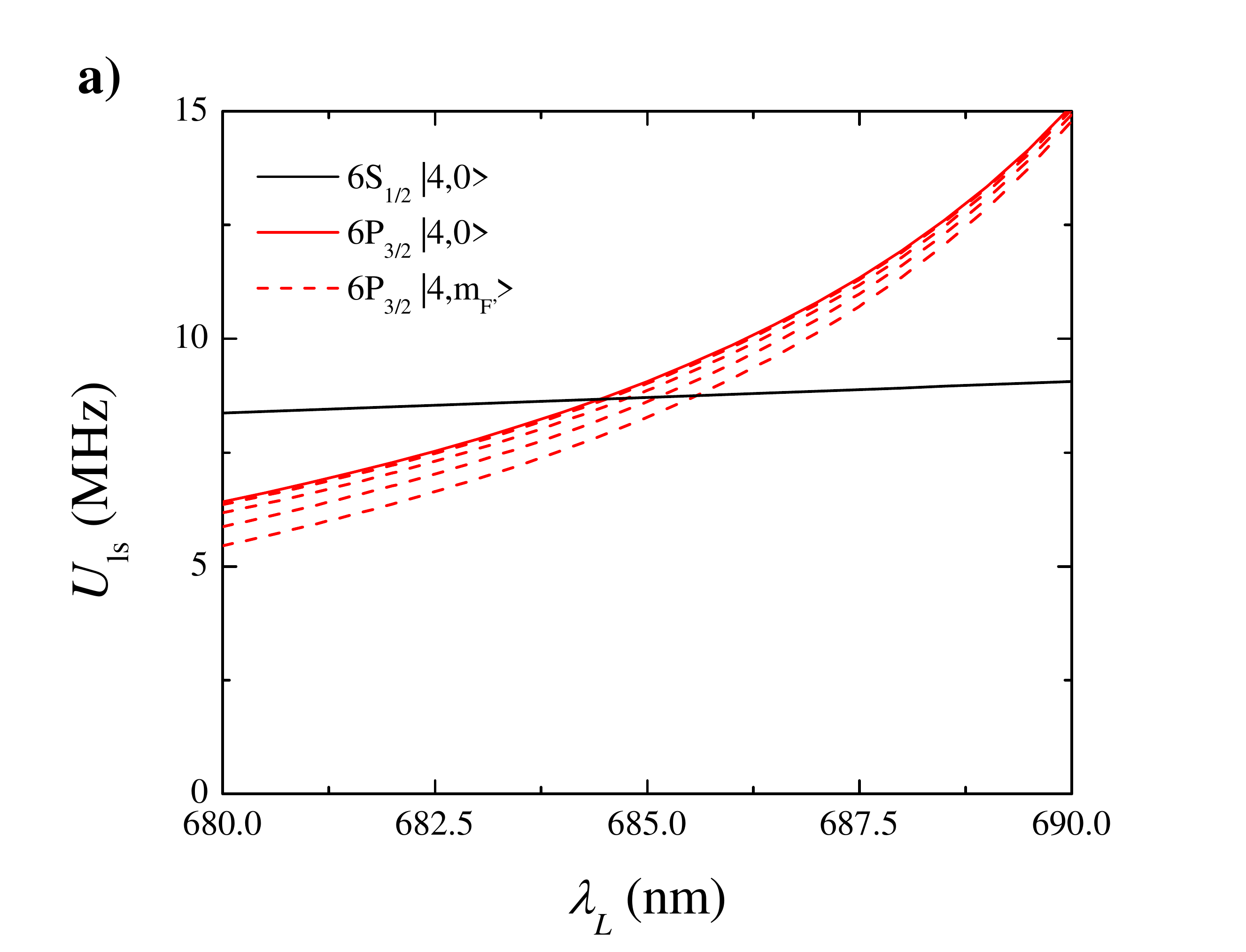}
\label{fig6:subfig2}\includegraphics[width=8cm]{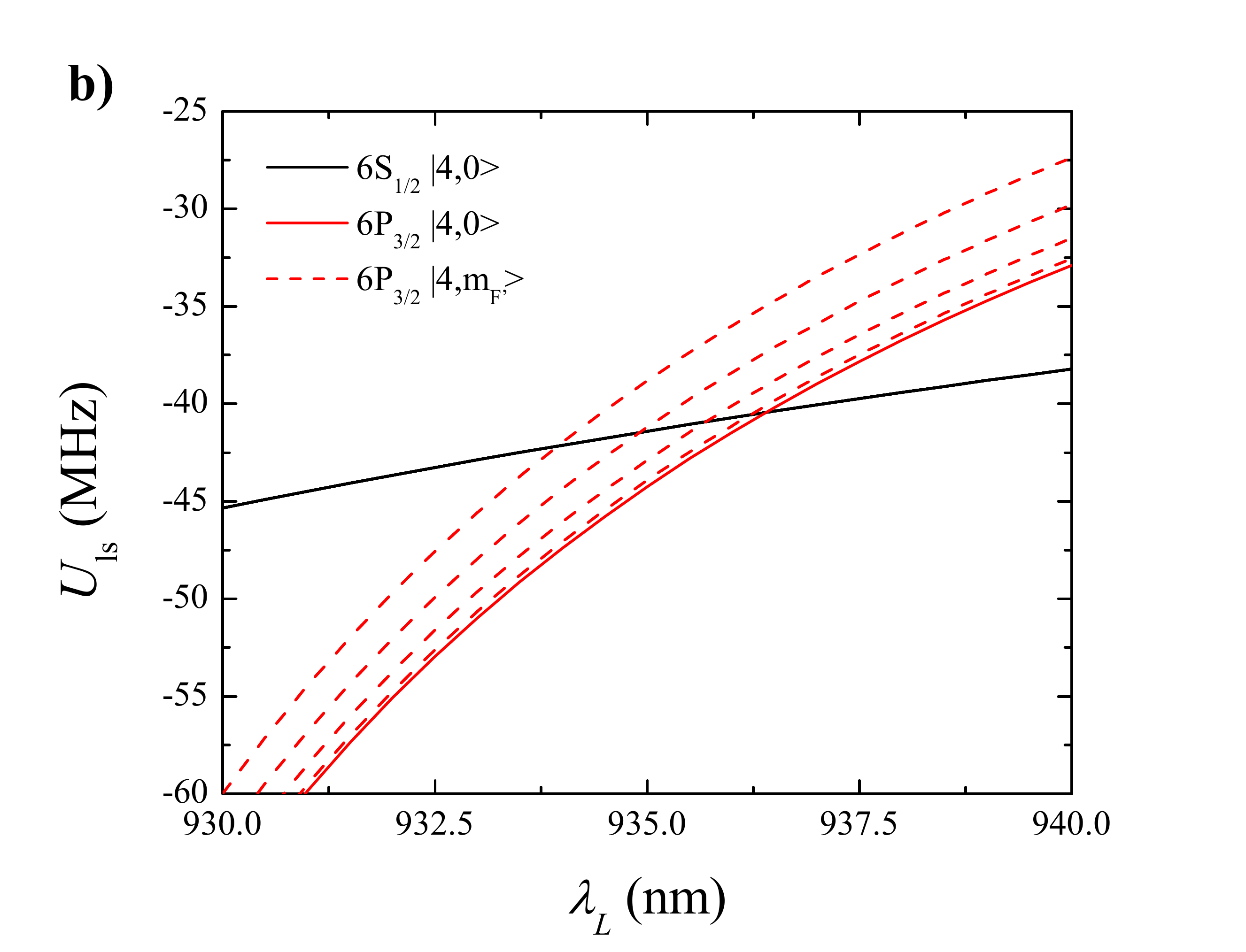}
\vspace{-2mm}
\caption{Replacement for Fig. 6 in Ref. \cite{lacroute2012} for determination of the magic wavelengths for the $6S_{1/2},F=4 \rightarrow 6P_{3/2},F'=4 $ transition of the Cs $D_2$ line. The light shifts $U_{ls}$ are for a linearly polarized beam with constant intensity $2.9\times10^9 ~\text{W m}^{-2}$ around (a) the blue-detuned magic wavelength at $\lambda_{\text{blue}}\simeq 684.9$ nm and (b) red-detuned magic wavelength at $\lambda_{\text{red}}\simeq 935.3$ nm.}
\label{fig6}
\end{figure*}

\begin{samepage}
\begin{figure*}[htbp]
\vspace{-5mm}
\centering
\label{fig7:subfig1}\includegraphics[width=7.5cm]{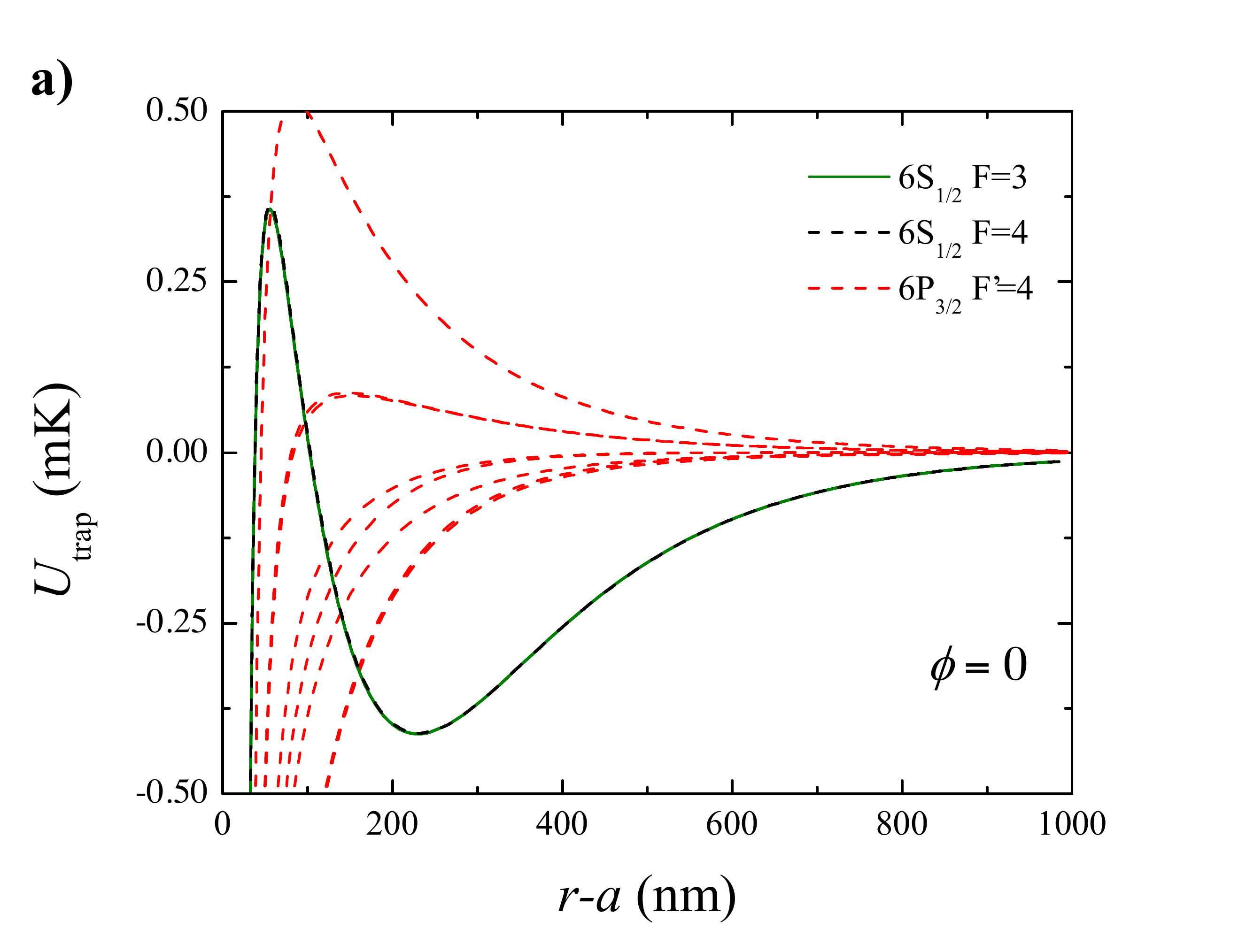}
\label{fig7:subfig2}\includegraphics[width=7.5cm]{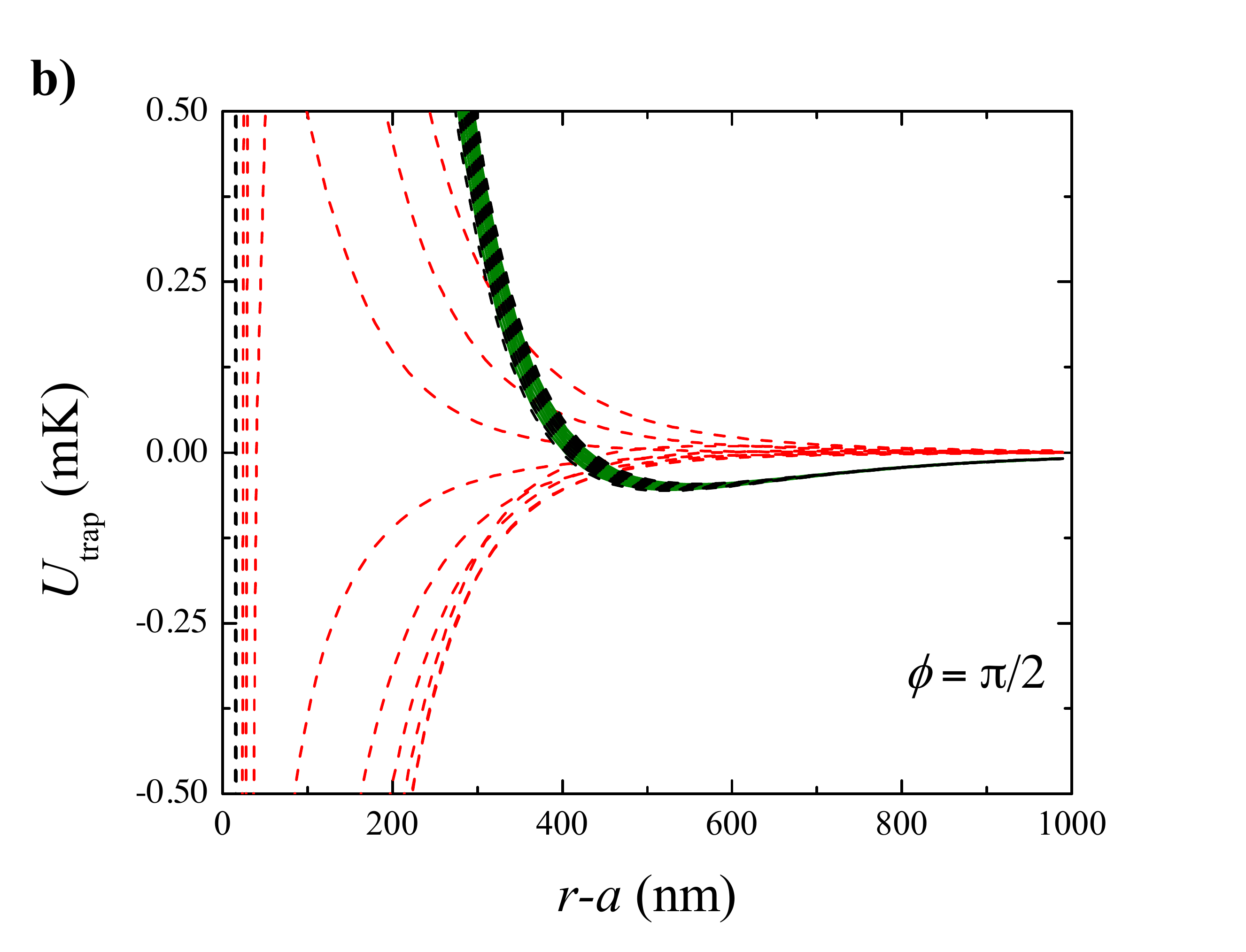}
\vspace{-2mm}
\caption{Replacement for Fig. 7 in Ref. \cite{lacroute2012}. Radial dependence of the trapping potential of the ground and excited states for the parameters used in Ref. \cite{vetsch2010} at $z = 0$. The polarization configuration is
the same as Fig. 1(b) of Ref. \cite{lacroute2012}. The energy sublevels of the ground states $F = 3$ and $F = 4$ of $6S_{1/2}$ are shown as solid green and dashed black curves, and the $F' = 4$ sublevels of the electronically excited state ($6P_{3/2}$) are shown as red dashed curves. (a)  Radial potential along $\phi=0$. The trap minimum is located at about 230 nm from the fiber surface. (b) Radial potential along $\phi=\pi/2$. }
\label{fig7}

\vspace{-3mm}

\centering
\label{fig8:subfig1}\includegraphics[width=7.5cm]{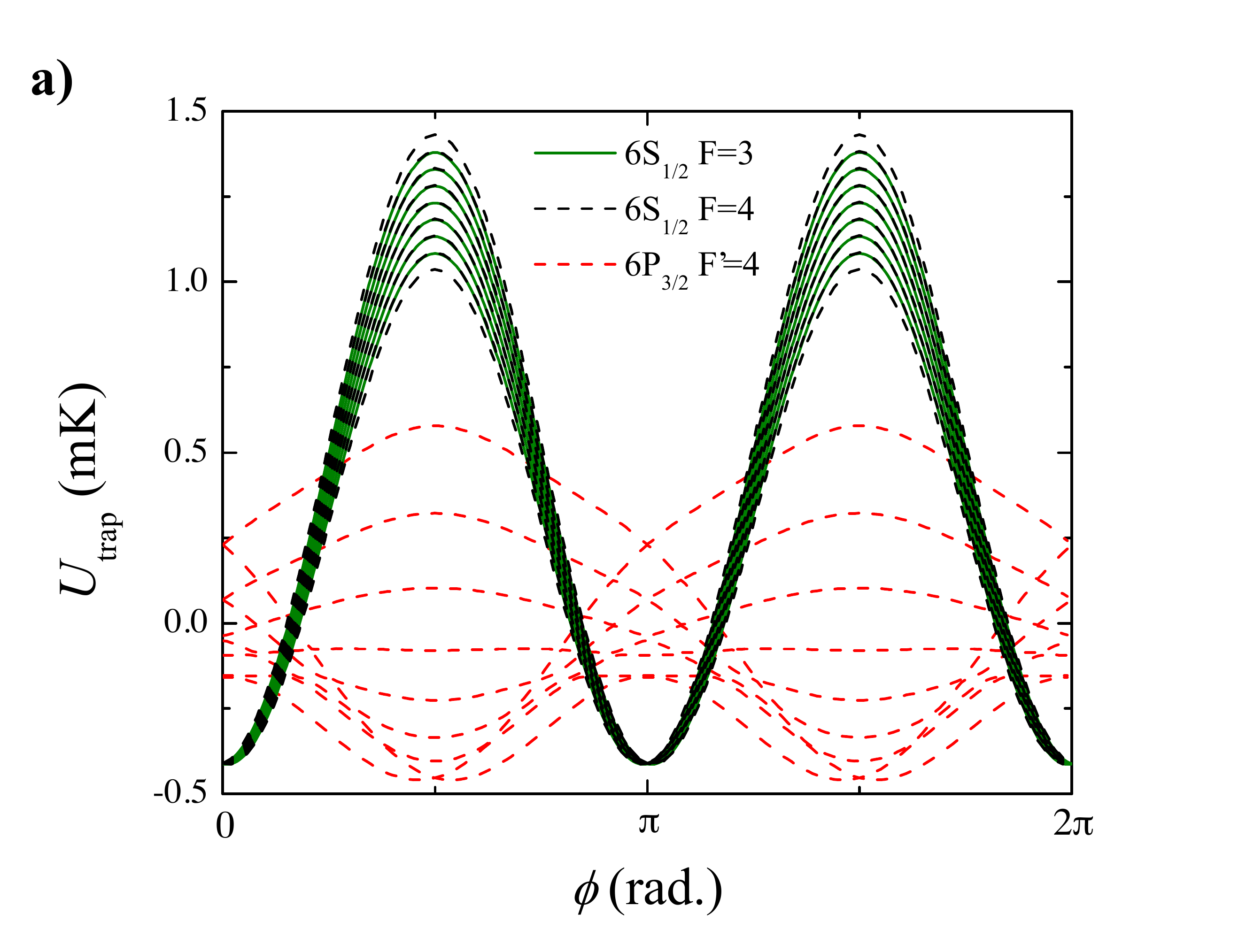}
\label{fig8:subfig2}\includegraphics[width=7.5cm]{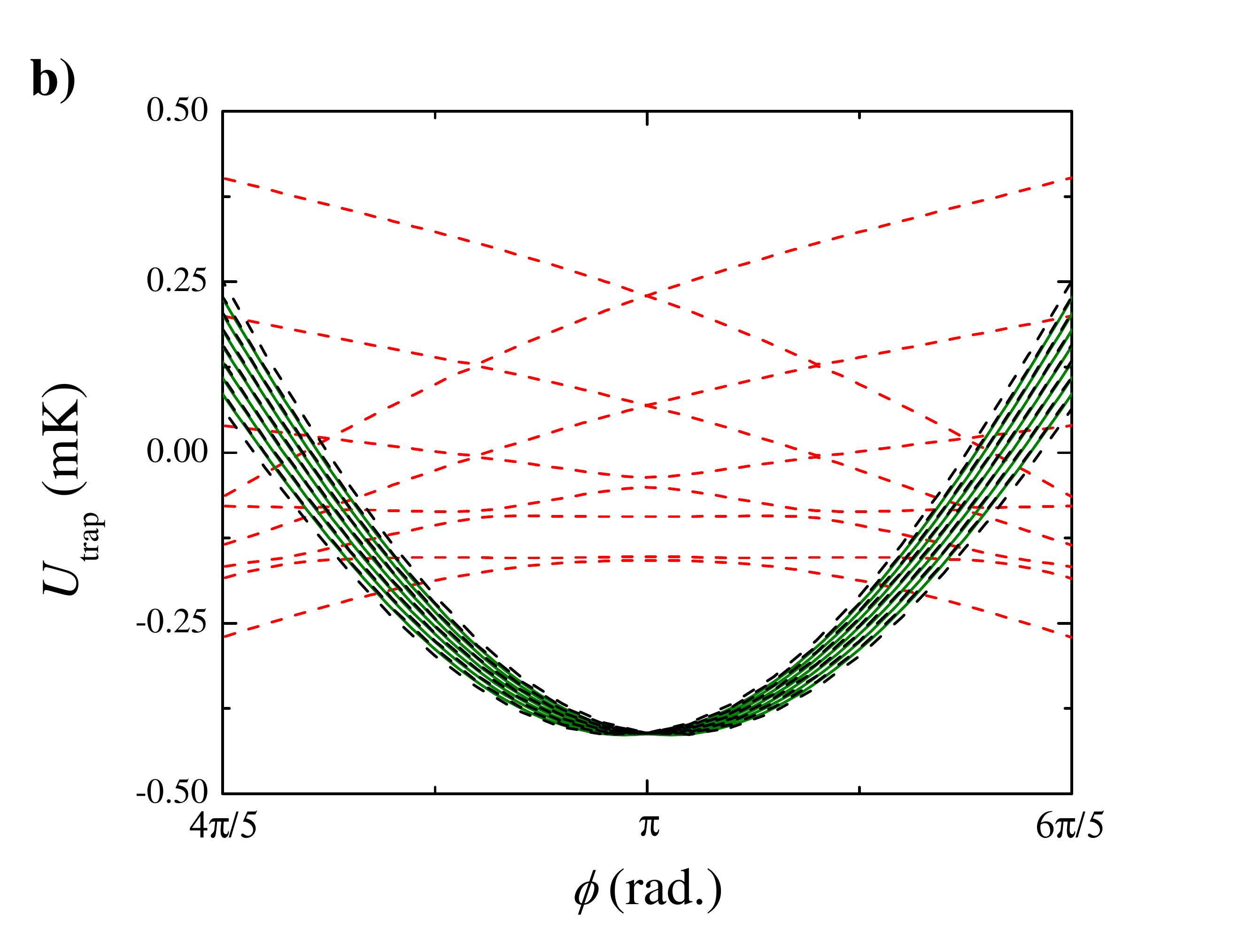}
\vspace{-2mm}
\caption{Replacement for Fig. 8 in Ref. \cite{lacroute2012} for the configuration of Ref. \cite{vetsch2010}. Azimuthal dependence of the trapping potential of the ground and excited states for the parameters used in Fig. \ref{fig7}. $r-a=230$ nm and $z = 0$. (a) The ground-state splitting is minimum
for $\phi=0$ and $\phi=\pi$. (b) Expanded view of (a) near a trap minimum at $\phi\simeq\pi$.}
\label{fig8} 

\vspace{-3mm}

\centering
\label{fig9:subfig1}\includegraphics[width=7.5cm]{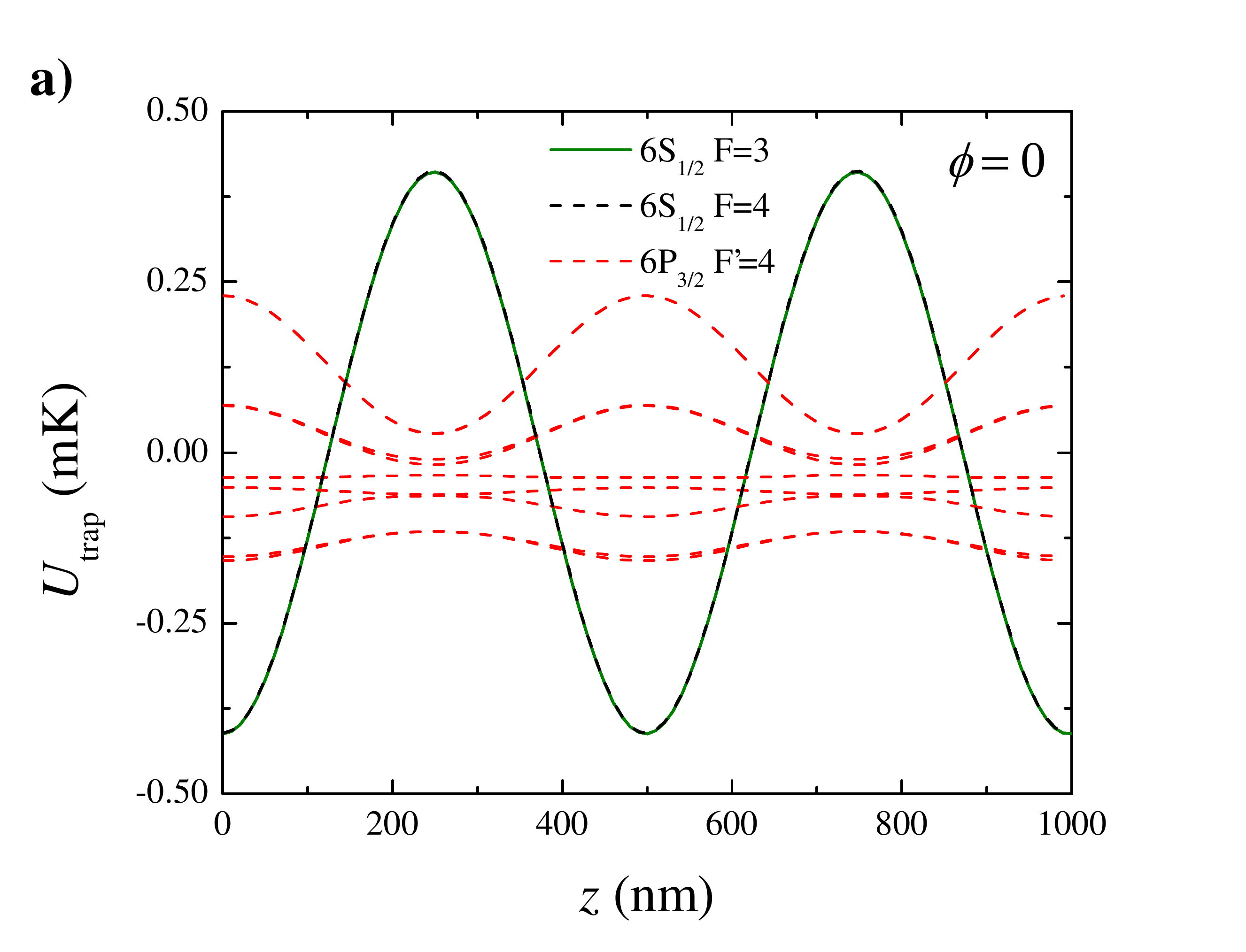}
\label{fig9:subfig2}\includegraphics[width=7.5cm]{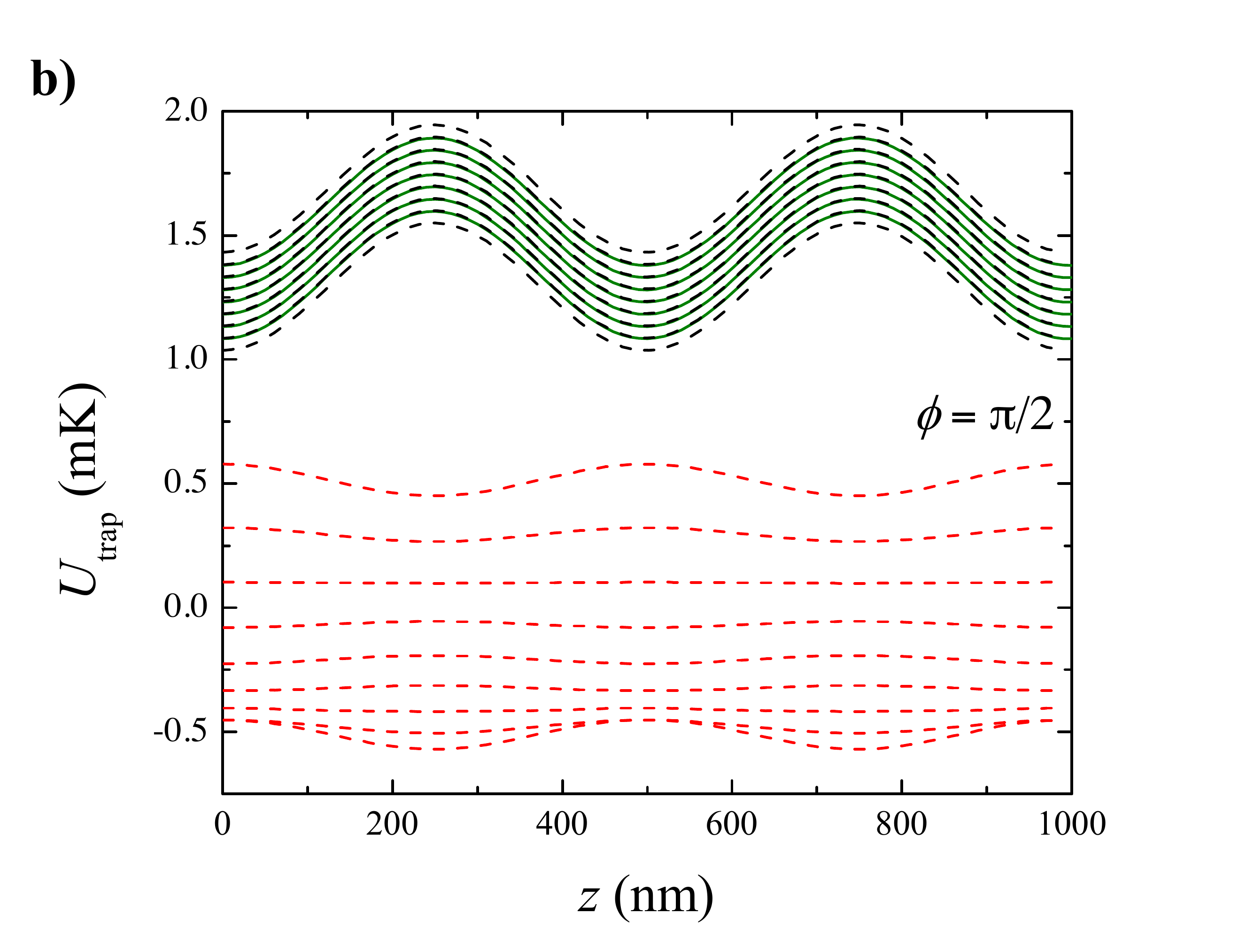}
\vspace{-2mm}
\caption{Replacement for Fig. 9 in Ref. \cite{lacroute2012} for the configuration of Ref. \cite{vetsch2010}. Axial dependence of the trapping potential for the ground and excited
states for the parameters used in Fig. \ref{fig7}.
(a) Longitudinal potential along $\phi= 0$. The distance from the fiber surface is
set to $r-a=230$ nm at the trap minimum. (b) Longitudinal potential along $\phi=\pi/2$. The distance from the fiber surface is again set to 230 nm.}
\label{fig9}
\end{figure*}
\end{samepage}

\begin{figure*}[htbp]
\centering
\label{fig10:subfig1}\includegraphics[width=8cm]{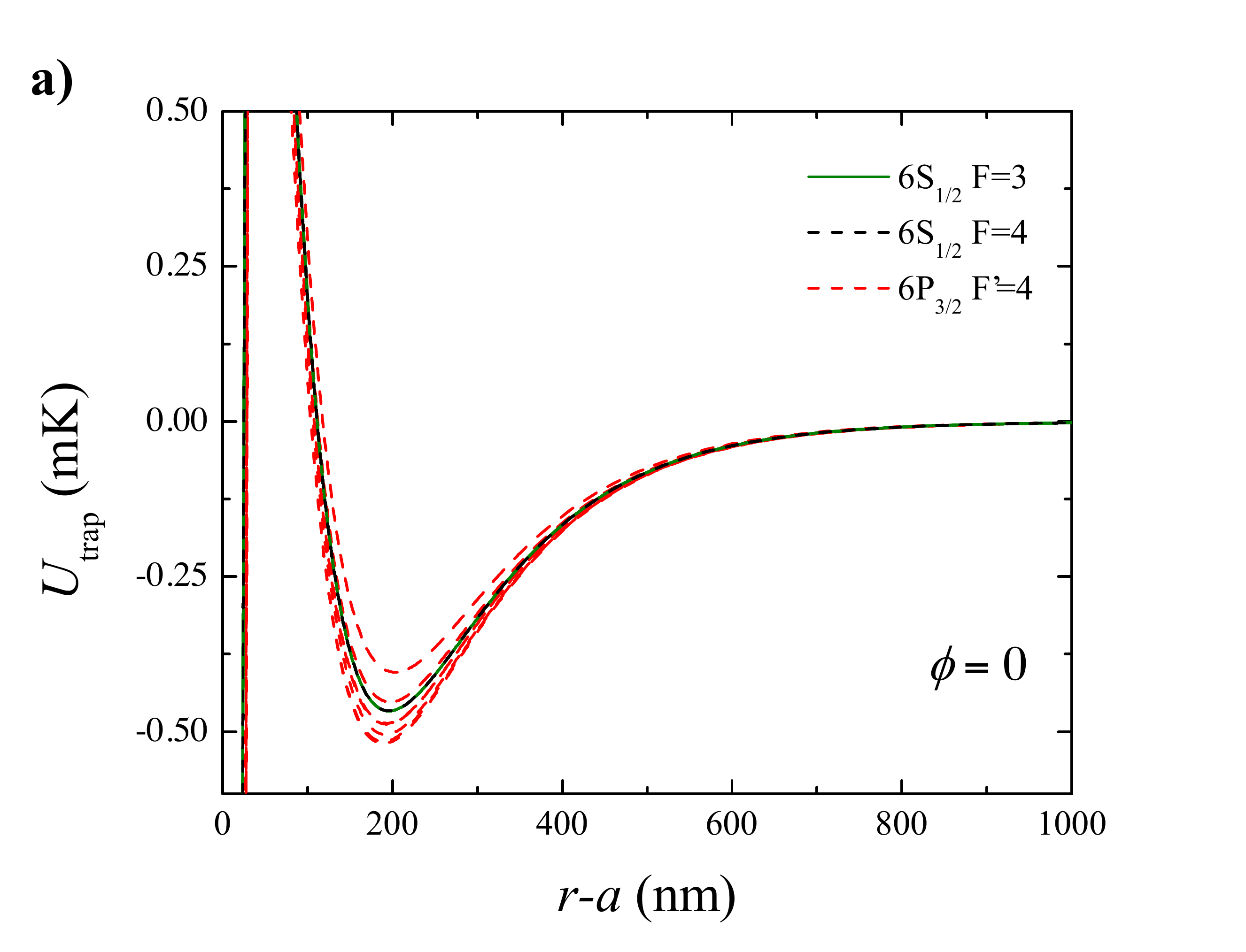}
\label{fig10:subfig2}\includegraphics[width=8cm]{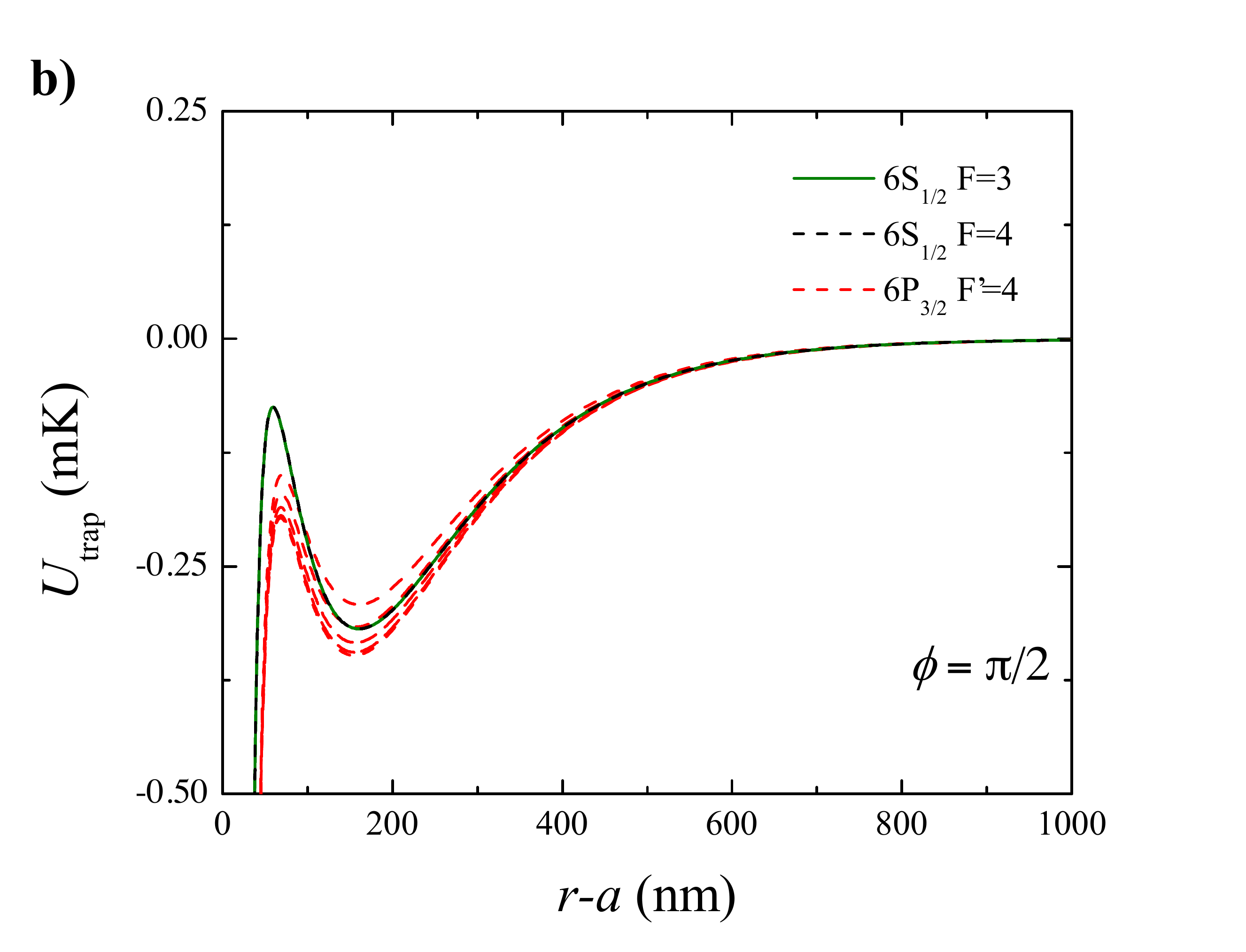}\\
\label{fig10:subfig3}\includegraphics[width=8cm]{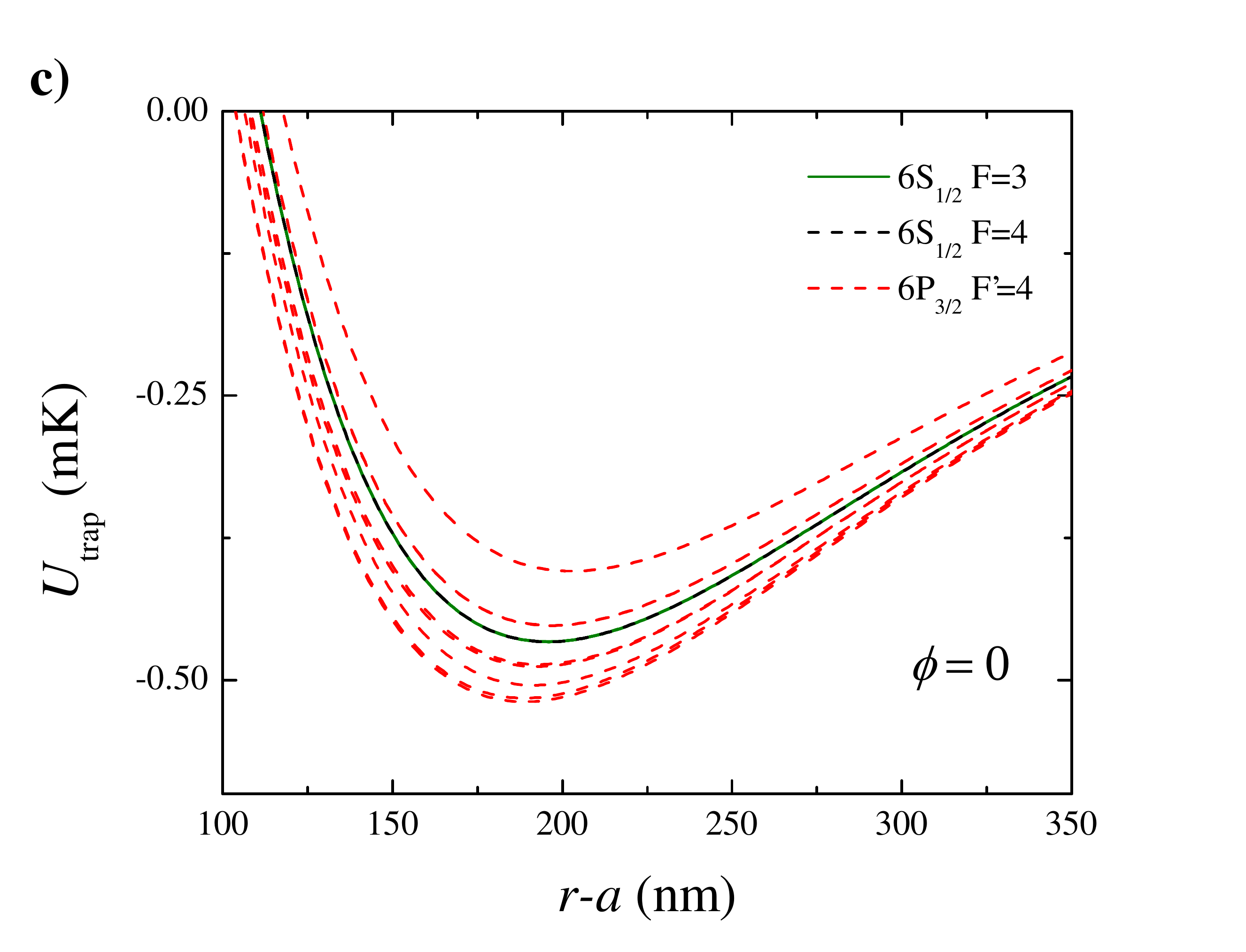}
\label{fig10:subfig4}\includegraphics[width=8cm]{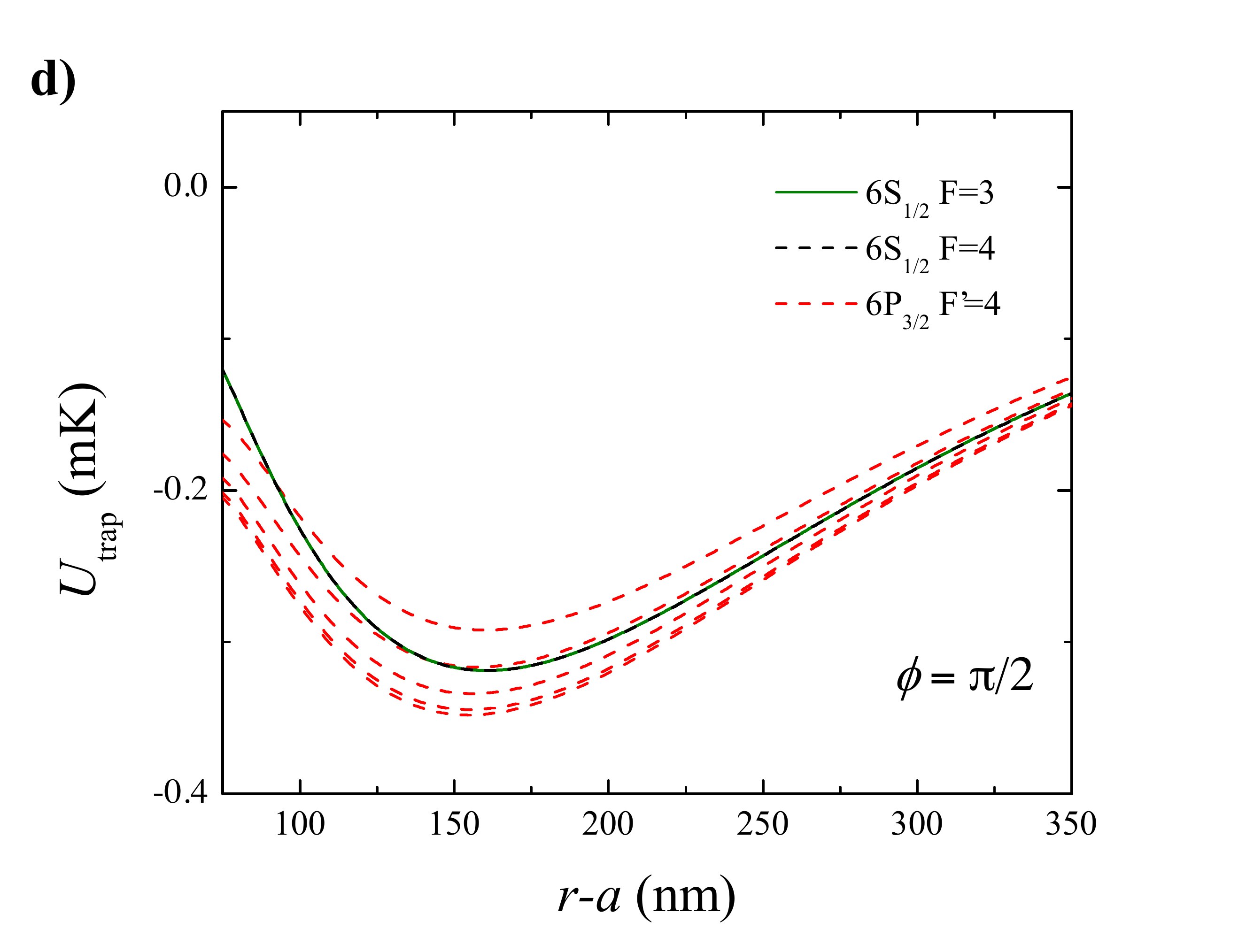}
\caption{Replacement for Fig. 10 in Ref. \cite{lacroute2012}. Radial dependence of the trapping potentials of the ground and
excited states using the magic wavelengths and the compensated configuration shown in
Fig. 1(d) of Ref. \cite{lacroute2012}. All beams are polarized along $\phi= 0$ (i.e., $\varphi_0 = 0$). The 935.3 nm beams
each have a power of 0.95 mW. The 684.9 nm beams each have a power of 16 mW. (a) Radial potentials along $\phi= 0$ (i.e., $\varphi_0 = 0$). The trap minimum for $6S_{1/2}$ is located at about 200 nm from the fiber surface. (b) Radial potential along $\phi=\pi/2$. (c), (d) Expanded views of (a) and (b) around the trap mininum.}
\label{fig10}
\end{figure*}

\begin{figure*}[htbp]
\centering
\label{fig11:subfig1}\includegraphics[width=8cm]{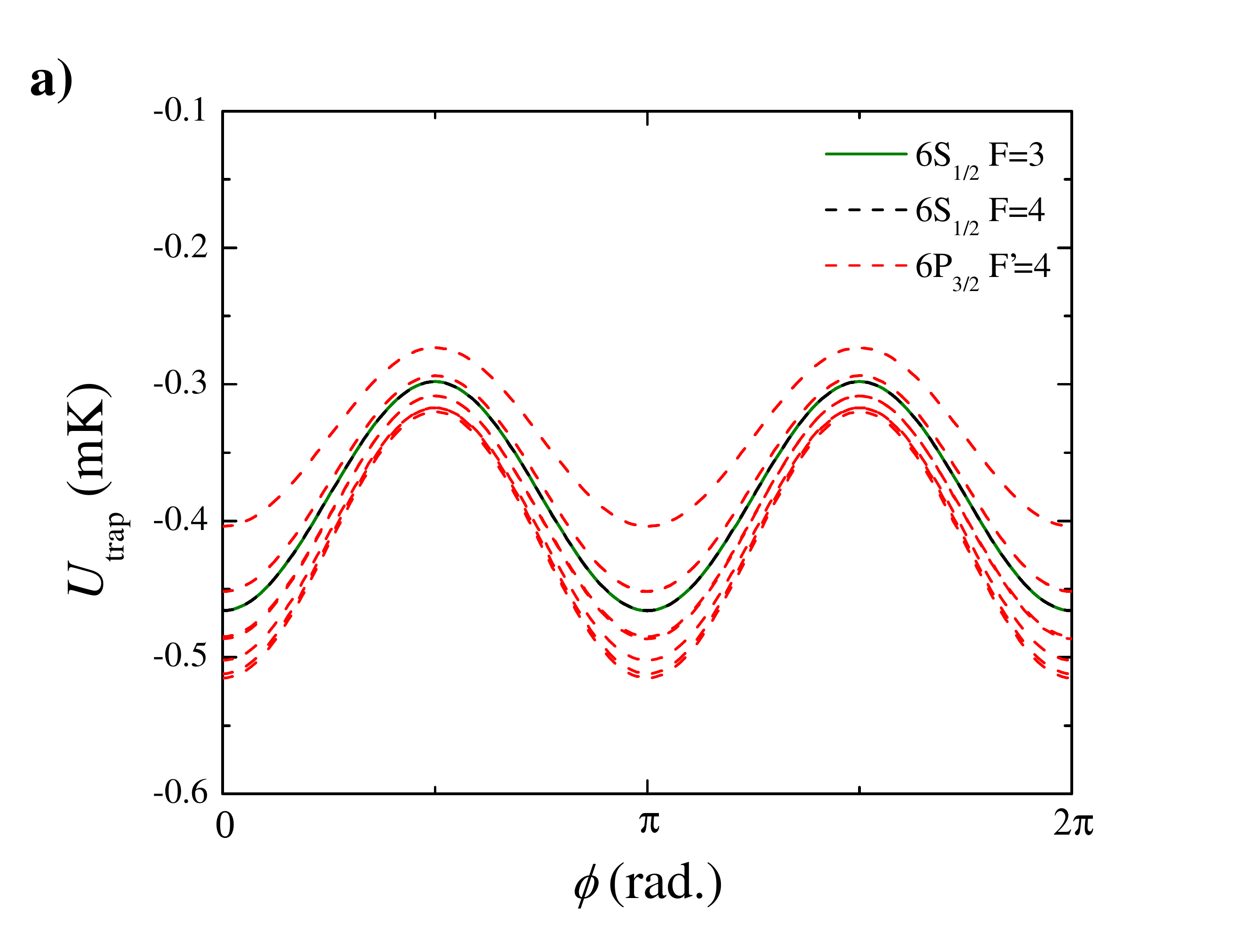}
\label{fig11:subfig2}\includegraphics[width=8cm]{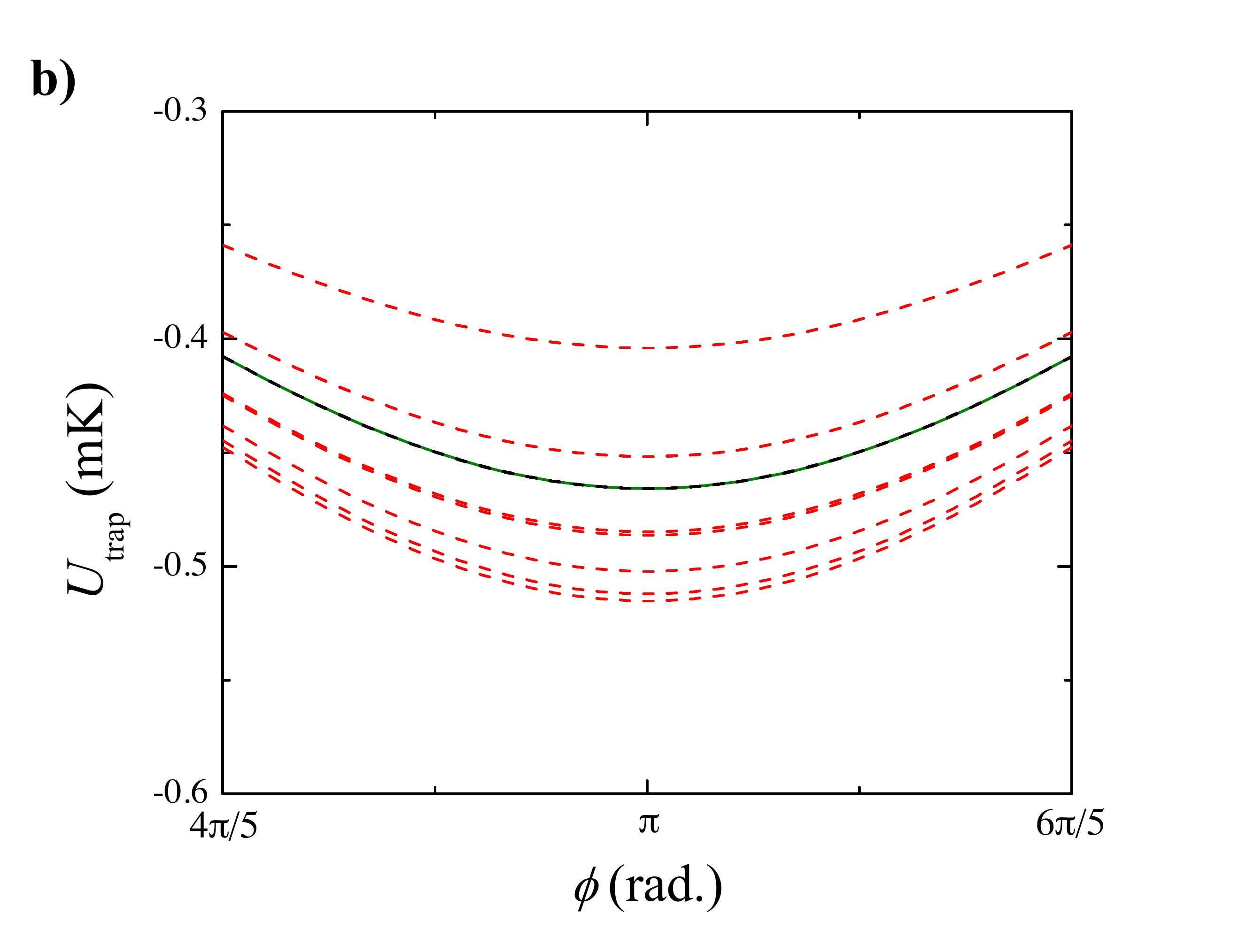}
\caption{Replacement for Fig. 11 in Ref. \cite{lacroute2012} for the magic, compensated scheme with $\lambda_{\text{red}} = 935.3$ nm and $\lambda_{\text{blue}} = 684.9$ nm. Azimuthal dependence of the trapping potential of the ground and
excited states for the parameters used in Fig. \ref{fig10}. $r-a = 200$ nm and $z = 0$. (b) Expanded view of (a) near a trap minimum at $\phi=\pi$.}
\label{fig11} 
\end{figure*}

\begin{figure*}[htbp]
\centering
\label{fig12:subfig1}\includegraphics[width=8cm]{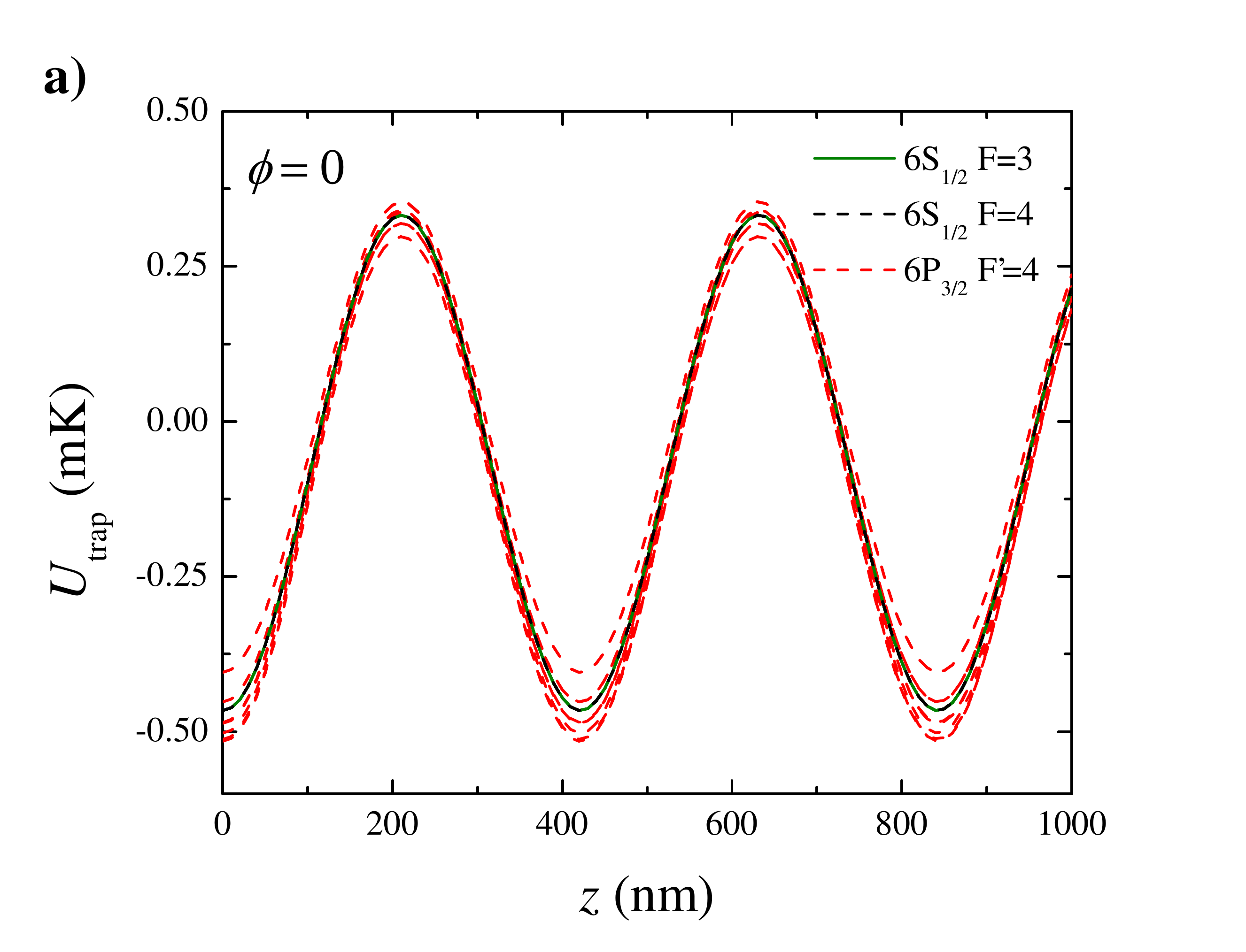}
\label{fig12:subfig2}\includegraphics[width=8cm]{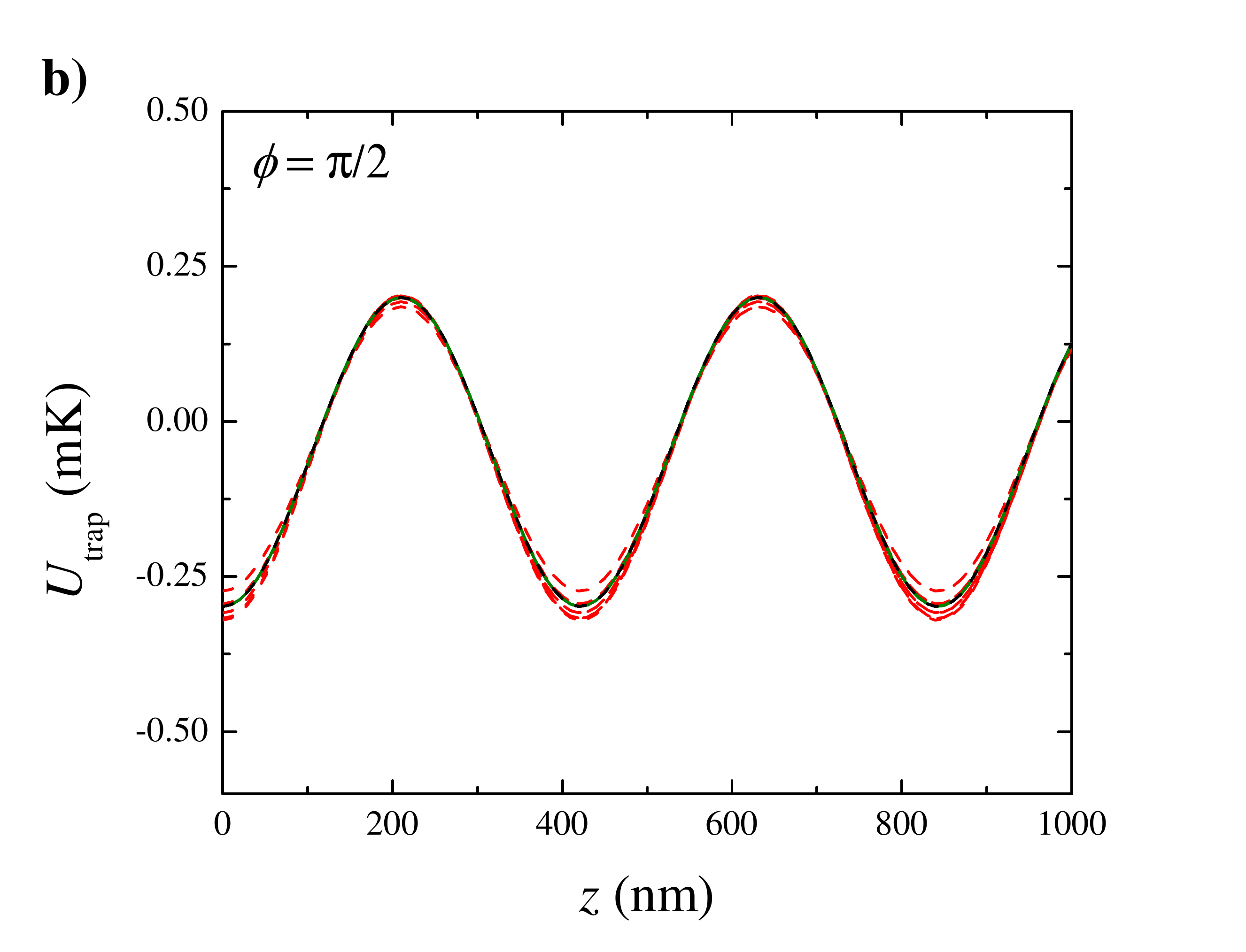}
\caption{Replacement for Fig. 12 in Ref. \cite{lacroute2012} for the magic, compensated scheme with $\lambda_{\text{red}} = 935.3$ nm and $\lambda_{\text{blue}} = 684.9$ nm. Axial dependence of the trapping potential for the ground and excited
states for the parameters used in Fig. \ref{fig10}. (a) Longitudinal potential along $\phi= 0$. The distance from the fiber surface is
set to $r-a = 200$ nm at the trap minimum. (b) Longitudinal potential along $\phi=\pi/2$. The distance from the fiber surface is again set to 200 nm.}
\label{fig12}
\end{figure*}

\newpage

\renewcommand{\thefigure}{\Alph{figure}\cite{goban2012}}
\setcounter{figure}{0}

\begin{figure*}[htbp]
\centering
\label{fig13:subfig2}\includegraphics[width=12cm]{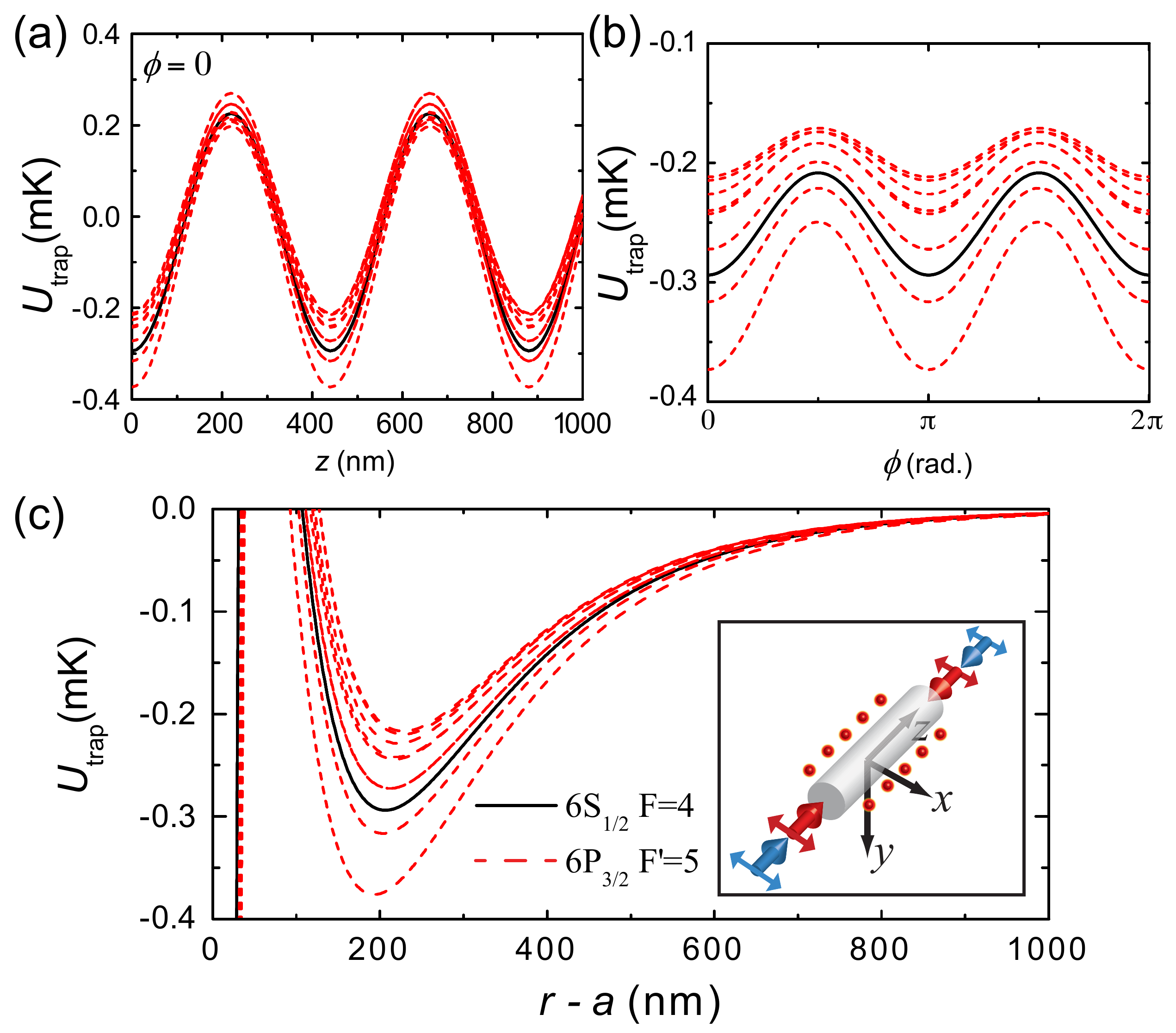}
\caption{The magic, compensated scheme with a pair of counter-propagating $x$-polarized ($\varphi_0=0$) red-detuned beams ($P_{\text{red}}=2\times0.4$ mW) at  $\lambda_{\text{red}} = 935.7$ nm and counter-propagating, $x$-polarized blue-detuned beams ($P_{\text{blue}} = 2\times5$ mW) at $\lambda_{\text{blue}} = 684.8$ nm. The distance from the fiber surface is set to 207 nm. (a) Azimuthal $U_{\rm trap}(\phi)$, (b) axial $U_{\rm trap}(z)$, (c) radial $U_{\rm trap}(r-a)$. Each black and red line corresponds to different energy eigenstates of the ground state ($6S_{1/2}, F=4$) and excited state ($6P_{3/2}, F'=5$), respectively.}
\label{fig13}
\end{figure*}

\newpage

\FloatBarrier

\begin{figure*}[htbp]
\label{fig14:subfig1}\includegraphics[width=9cm]{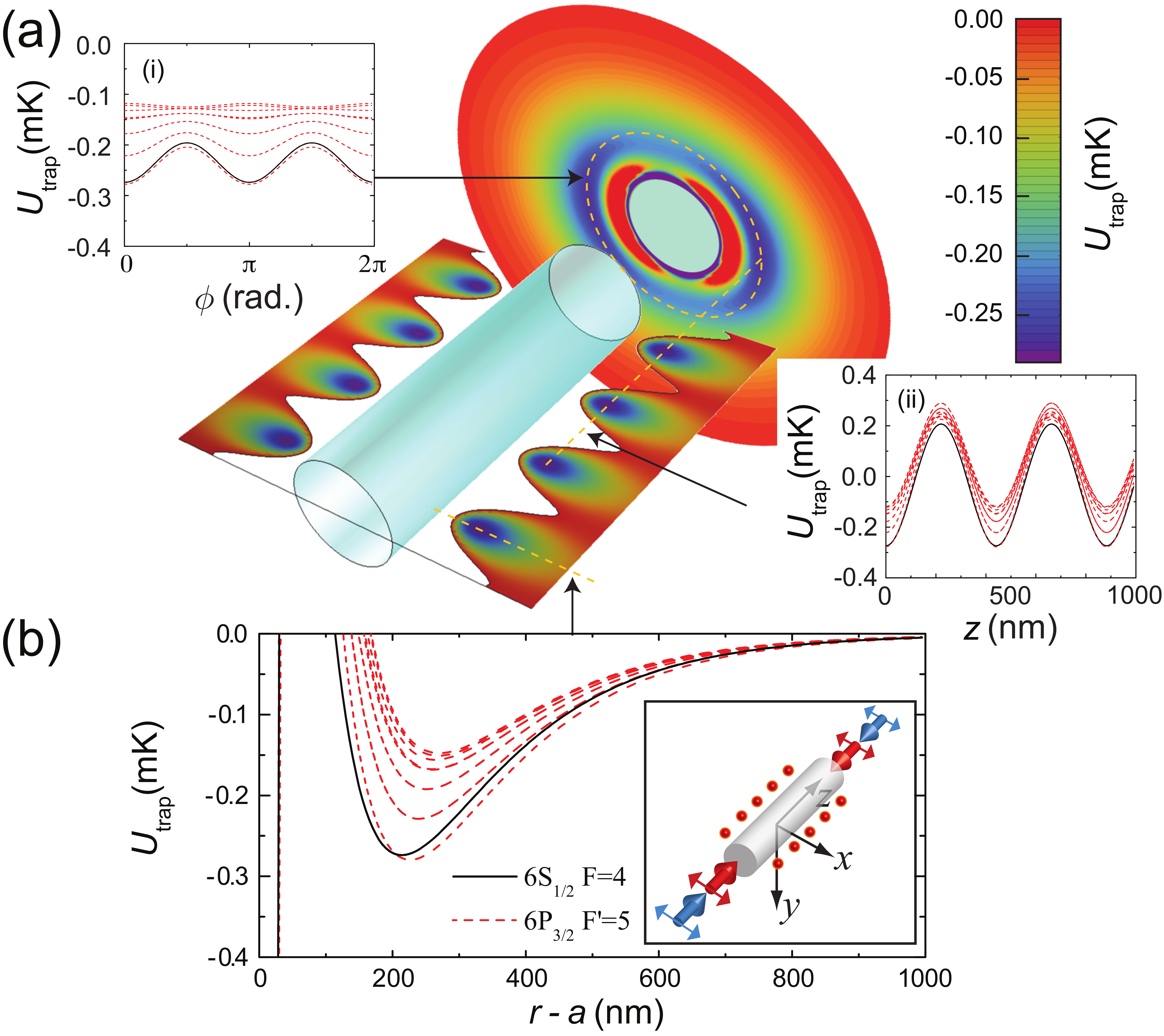}
\caption{Replacement for Fig. 1 in Ref. \cite{goban2012} for the actual wavelengths used in our experiment \cite{goban2012}. We used a pair of counter-propagating $x$-polarized ($\varphi_0=0$) red-detuned beams ($P_{\text{red}}=2\times0.4$ mW) at  $\lambda_{\text{red}} = 937.1$ nm, and counter-propagating, $x$-polarized blue-detuned beams ($P_{\text{blue}} = 2\times5$ mW) at $\lambda_{\text{blue}} = 686.1$ nm in our experiment \cite{goban2012}. The resulting interference is averaged out by detuning the beams to $\delta_{fb} = 382$ GHz. Adiabatic trapping potential $U_{\text{trap}}$ for a state-insensitive, compensated nanofiber trap for the $6S_{1/2}, F=4$ states in atomic Cs outside of a cylindrical waveguide of radius $a=215$ nm. $U_{\text{trap}}$ for the substates of the ground level $F=4$ of $6S_{1/2}$ (excited level $F'=5$ of $6P_{3/2}$) are shown as black (red-dashed) curves. \textbf{(a)(i)} azimuthal $U_{\text{trap}}(\phi)$, \textbf{(ii)} axial $U_{\text{trap}}(z)$ and \textbf{(b)} radial $U_{\text{trap}}(r-a)$ trapping potentials. The trap minimum for $6S_{1/2}$ is located at about 215 nm from the fiber surface. Input polarizations for the trapping beams are denoted by the red and blue arrows in the inset in (b).}
\label{fig14}
\end{figure*}

\begin{figure*}[htbp]
\centering
\label{fig15:subfig2}\includegraphics[width=9cm]{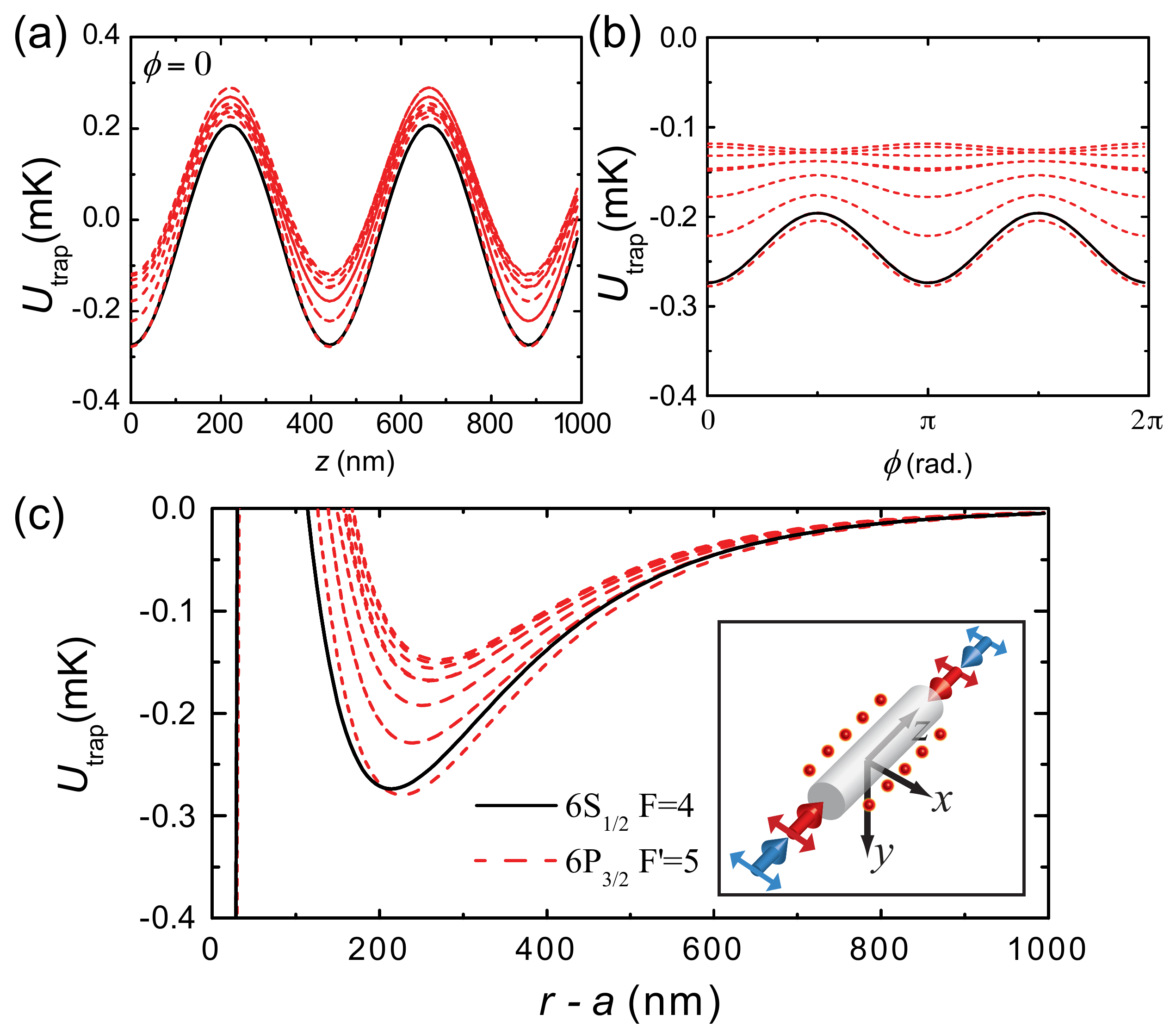}
\caption{Replacement for Fig. SM5 in the supplemental material of Ref. \cite{goban2012} for actual wavelengths used in our experiment \cite{goban2012}. We used a pair of counter-propagating $x$-polarized ($\varphi_0=0$) red-detuned beams ($P_{\text{red}}=2\times0.4$ mW) at  $\lambda_{\text{red}} = 937.1$ nm, and counter-propagating, $x$-polarized blue-detuned beams ($P_{\text{blue}} = 2\times5$ mW) at $\lambda_{\text{blue}} = 686.1$ nm in our experiment \cite{goban2012}. Expanded view of the insets in Fig. \ref{fig14}. The distance from the fiber surface is set to 215 nm. (a) Azimuthal $U_{\rm trap}(\phi)$, (b) axial $U_{\rm trap}(z)$, (c) radial $U_{\rm trap}(r-a)$. Each black and red line corresponds to different energy eigenstates of the ground state ($6S_{1/2}, F=4$) and excited state ($6P_{3/2}, F'=5$), respectively.}
\label{fig15}
\end{figure*}

\end{document}